\documentclass[sigconf]{acmart}

\usepackage[english]{babel}
\usepackage{blindtext}
\usepackage{import}
\usepackage{graphicx}
\graphicspath{ {images/} }
\usepackage{listings}
\usepackage{caption}
\usepackage{subcaption}
\usepackage{tabularx}
\usepackage{listings}
\usepackage{xcolor}
\usepackage{enumitem}
\setlist[itemize]{topsep=0pt, partopsep=0pt, parsep=0pt, itemsep=0pt, leftmargin=*}

\renewcommand\footnotetextcopyrightpermission[1]{} 
\setcopyright{none}

\acmDOI{ }

\acmISBN{ }

\acmConference[]{}
\acmYear{ }
\copyrightyear{ }

\acmPrice{}

\begin{document}

\title{Examining DOM Coordinate Effectiveness For Page Segmentation}




\author{Jason Carpenter, Faaiq Bilal, Eman Ramadan, Zhi-Li Zhang}
\email{{carpe415, bilal021, eman}@umn.edu, zhzhang@cs.umn.edu}
\affiliation{%
  \institution{University of Minnesota}
  \city{Minneapolis}
  \state{MN}
  \country{USA}
}

\renewcommand{\shortauthors}{Carpenter}

\begin{abstract}
Web pages form a cornerstone of available data for daily human consumption and with the rise of LLM-based search and learning systems a treasure trove of valuable data. The scale of this data and its unstructured format still continue to grow requiring ever more robust automated extraction and retrieval mechanisms. Existing work, leveraging the web pages Document Object Model (DOM), often derives clustering vectors from coordinates informed by the DOM such as visual placement or tree structure. The construction and component value of these vectors often go unexamined. Our work proposes and examines DOM coordinates in a detail to understand their impact on web page segmentation. Our work finds that there is no one-size-fits-all vector, and that visual coordinates under-perform compared to DOM coordinates by about 20-30\% on average. This challenges the necessity of including visual coordinates in clustering vectors. Further, our work finds that simple vectors, comprised of single coordinates, fare better than complex vectors constituting 68.2\% of the top performing vectors of the pages examined. Finally, we find that if a vector, clustering algorithm, and page are properly matched, one can achieve overall high segmentation accuracy at 74\%. This constitutes a 20\% improvement over a naive application of vectors. Conclusively, our results challenge the current orthodoxy for segmentation vector creation, opens up the possibility to optimize page segmentation via clustering on DOM coordinates, and highlights the importance of finding mechanisms to match the best approach for web page segmentation.
\end{abstract}


\maketitle

\section{Introduction}
Access to information defines this century. Vast sums of data are collected, exchanged, and analyzed across the Internet. Search engines, by providing access to this information on web pages, revolutionized the world. Further, Large Language Models (LLMs) have revolutionized access to information by leveraging knowledge from extensive web-scale datasets. From a web page perspective, the Internet is estimated to have between 1.2 and 40 billion web pages\cite{size-of-internet-webpages, size-of-internet-webpages-2, size-of-the-internet-webpages-3}. Important information is still widely published on websites such as Amazon, eBay, Craigslist, Twitter, Facebook, Instagram, etc. Simply put, web pages form the backbone of the Internet. Continued access to this data is crucial for the development of information-driven technologies such as LLMs, search engines, and aggregation services. However, due to the diverse and unstructured nature of web pages, information extraction through automated means is difficult~\cite{dom-extraction-6-the-good-one, webpage-segmentation-1}. Older approaches using web scraping technologies are fragile and require manual interventions to handle changing webpage layouts~\cite{web-page-scraping-1}. Some works attempt to utilize machine learning and visual queues, but are often limited by the cost of machine learning. 

\begin{figure}
    \centering
    \includegraphics[width=0.7\linewidth]{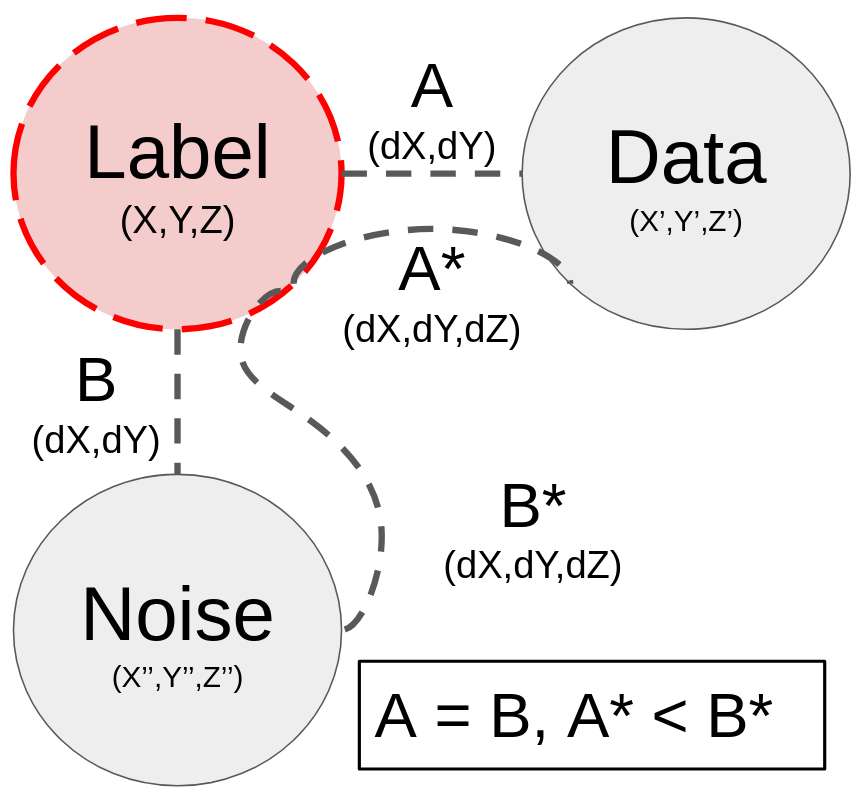}
    \vspace{-0.1in}
    \caption{New Coordinates Required to Break Ambiguity}
    \label{fig:distance-ambiguity}
\end{figure}

Page segmentation requires the creation of useful associations between direct HTML Document Object Model (DOM) nodes in a web page~\cite{webpage-segmentation-4}. Such associations provide value for grouping similar items, attaching labels, or dividing concepts, such as news articles. In clustering terms, this would be by decreasing the ``distance'' between nodes that should be associated. Existing page segmentation works frequently use metrics derived from the DOM~\cite{webpage-segmentation-1}. However, many of these works do not examine or evaluate the effectiveness of vector compositions derived from the DOM. In some cases, aspects of such vectors, such as visual placement coordinates, are used with little verification\cite{webpage-segmentation-1}. For example, an intuitive connection is visible distance, such as the pixel distance between items on a page. This distance, while intuitive, is not always correct as shown in Fig.~\ref{fig:distance-ambiguity}. 
A robust exploration of how DOM-informed coordinates is therefore valuable to better understand the impact of such selections.
Moreover, new coordinates are required to resolve the ambiguity in the association between nodes that are otherwise conceptually associated.
This concept can be broadly expanded to accommodate any coordinates selected for such a distance measure, as even DOM coordinates may fail to associate the related fields properly.

In this paper, we propose and examine the effectiveness of DOM-informed coordinates for web page segmentation.
This paper's contributions are:
\begin{itemize}
\item We propose and evaluate new DOM coordinates: Tag Depth, Div ID, Tag Group, and Data Index, informed by our survey of web pages and clustering outcomes.
    \item We conduct a robust component-wise evaluation of coordinate vectors composed of new and existing DOM coordinates to understand, in detail, their capacity for page segmentation.
    
\end{itemize}
Our insights are summarized as follows:
\begin{itemize}
    \item Comparing single-coordinate vectors and multiple-coordinate vectors, we find that single coordinate vectors were the best performers in 68.2\% of the tested web pages. However, we find a small number of web pages cluster very well with multiple-coordinate vectors.
    \item Cartesian coordinates are a cornerstone of existing page segmentation vectors. Surprisingly, we find that structural DOM coordinates, such as Div IDs, perform roughly twice as well as Cartesian coordinates. 
    \item We examine the performance of HDBSCAN and OPTICS on these new coordinates and find that OPTICS achieves 2\% better performance than HDBSCAN. However, the sets of web pages on which each algorithm performs well are slightly different, creating a complementary optimization. Notably, this results in an average 20\% improvement to the overall Rand score when the optimal coordinates are selected.
    \item Given the diversity of web pages, we find that different vectors and algorithms perform better on different sets of web pages. If the best individual performer is selected, we see an average Rand score of 54\% across the dataset. However, if we pair the best coordinate vector and algorithm for each web page, the average Rand score rises to 74\%.
\end{itemize}
From these insights, we find that the existing understanding of how segmentation vectors are constructed is not entirely settled, and there is still room for improvement. Furthermore, the contributions of this work can be used to advance automatic page extraction through systems such as LLMs, web page scrapers, and more advanced reasoning systems, such as the semantic web.


\section{Challenges \& Related Work}
\label{sec:webpage-survey}
    \begin{figure}
    \centering
    \includegraphics[width=0.26\linewidth]{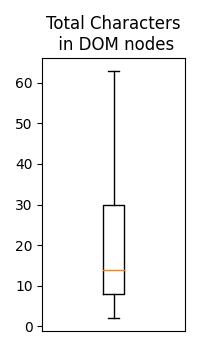}
    \includegraphics[width=0.27\linewidth]{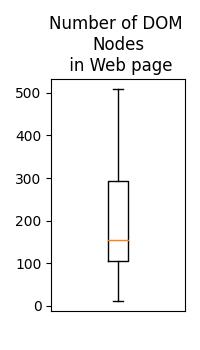}
    \includegraphics[width=0.27\linewidth]{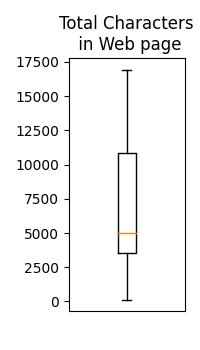}
    \vspace{-0.1in}
    \caption{DOM/text Analysis of Surveyed Web Pages}
    \label{fig:web-survey}
\end{figure}

Web pages are extremely diverse in terms of content and layout. Even within a single domain, web pages can differ fundamentally, for example, in how they lay out items horizontally or vertically. For our work, we conducted a small survey of web page metrics, including total DOM size, the number of textual characters in DOM nodes, and total textual characters in the web page. The examined web pages are described in Sec~\ref{dataset-creation}. The results outlined in Fig.~\ref{fig:web-survey} show that while the text count per DOM is relatively constrained, with about 75\% of the DOM nodes sitting between 1-30 characters, the total text and DOM nodes vary widely. 
This diversity presents a challenge as any coordinates collected may have a different expression in the pages themselves and thus complicate distance calculations for relation. 
For example, if a table of data is oriented horizontally with a format of ``data-label'', and a row below it with the same format, traditional two-dimensional visual distance calculations will incorrectly associate the two labels together and the two data items separately. Any examination and improvement to using clustering using web page coordinates must understand this specific coordinate interplay that may appear in pages.







Broadly, existing page segmentation uses aspects of the visual or DOM structure of a web page to group relevant parts of a page together~\cite{webpage-segmentation-2, webpage-clustering-2, webpage-segmentation-3, dom-extraction-6-the-good-one}. These mechanisms, while effective, often rely on assumptions that are not fully validated, such as in the case of visually informed coordinates~\cite{webpage-segmentation-1}. Some aspects of these coordinates may not hold up under scrutiny, such as the assumption in~\cite{webpage-segmentation-4} that related groups share font and color traits, or in \cite{webpage-segmentation-2}, such that the tag character of sub-trees in a DOM is necessarily related. Our work seeks to step back and approach such validation for DOM-informed and visual coordinates through a more careful, piece-wise evaluation.

\section{Methodology}
\label{dataset-creation}
We validate the effectiveness of DOM coordinates by proposing new coordinates and evaluating them against visual coordinates.
\begin{figure}
    \centering
    \includegraphics[width=0.5\textwidth]{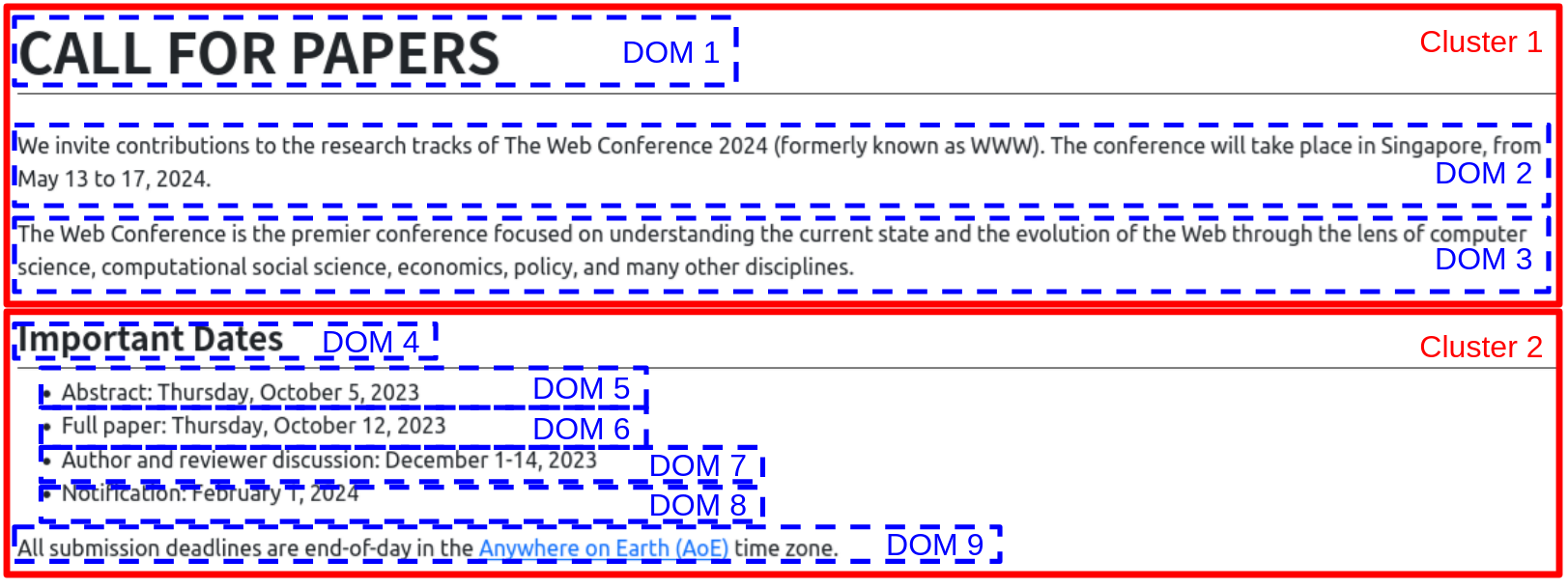}
    \vspace{-0.2in}
    \caption{Example of Clustering Approach for Annotations}
    \label{fig:sectional-annotation}
    \vspace{-0.25in}
\end{figure}
We surveyed 44 websites and collected a total of 23410 DOM nodes across all the web pages. We selected web pages from commonly visited websites and the Alexa top 1m dataset~\cite{dataset-alexa-top-1m, dataset-top-webpages}, and manually annotated conceptually related clusters. We used a broad definition for our related segments, as semantic similarity is loosely defined. We include an illustrated example of the annotation in Fig.~\ref{fig:sectional-annotation}, in which we select based on a conceptual similarity, the ground truth for how clusters should be formed. 

We define our new coordinates as follows: 
\begin{itemize}
\item \textbf{Tag Depth (Td):} quantifies the hierarchical depth of a DOM element in a document's structure, incrementing with each nested div, and decrementing when the corresponding tag is closed.
\item \textbf{Div ID (DI):} represents the linear position of an element within the DOM tree. It is a sequential identifier that distinguishes elements that are visually close but hierarchically distant.
\item \textbf{Tag Group (TG):} extends the concept of Td, incrementing for elements within the same sibling group. The value resets when a DOM branch ends.
\item \textbf{Data Index (DiD):} Similar to DI, this metric represents the linear position of elements in the DOM tree. However, DiD only increases for DOM elements containing text. 
\item \textbf{True-Cartesian (TX-TY):} uses the raw Cartesian values, which are likely offset, and calculates the true center of the DOM item with the height and width. We include this as an additional baseline to compliment traditional cartesian coordinates.
\end{itemize}
From these proposed coordinates, we create standalone (rows 1-4) and variable composition vectors as shown in (rows 5-13) in Table~\ref{tab:results-from-experiments}. 

To evaluate the effectiveness of each vector, we use HDBSCAN\cite{HDBSCAN} and OPTICS\cite{OPTICS} for clustering. We selected these two algorithms because: (1)~our clusters are characterized by variable sizes and density, for which our selected algorithms are suited~\cite{python-clustering-guide}. (2)~OPTICS has been used in other related work for webpage segmentation~\cite{webpage-segmentation-4}. 
We have three metrics for evaluation: 
\begin{itemize}
    \item \textbf{Rand Score:} A similarity score between two clusterings, specifically the algorithm's output and ground truth.
    \item \textbf{Cluster Count Difference:} Difference in the number of clusters between the algorithm's output and ground truth.
    \item \textbf{Cluster Size Difference:} Difference in the size of clusters from the algorithm's output and ground truth.
\end{itemize}

\section{Evaluation}
\label{sec:eval}
\begin{figure}
    \centering
    \includegraphics[width=0.21\textwidth]{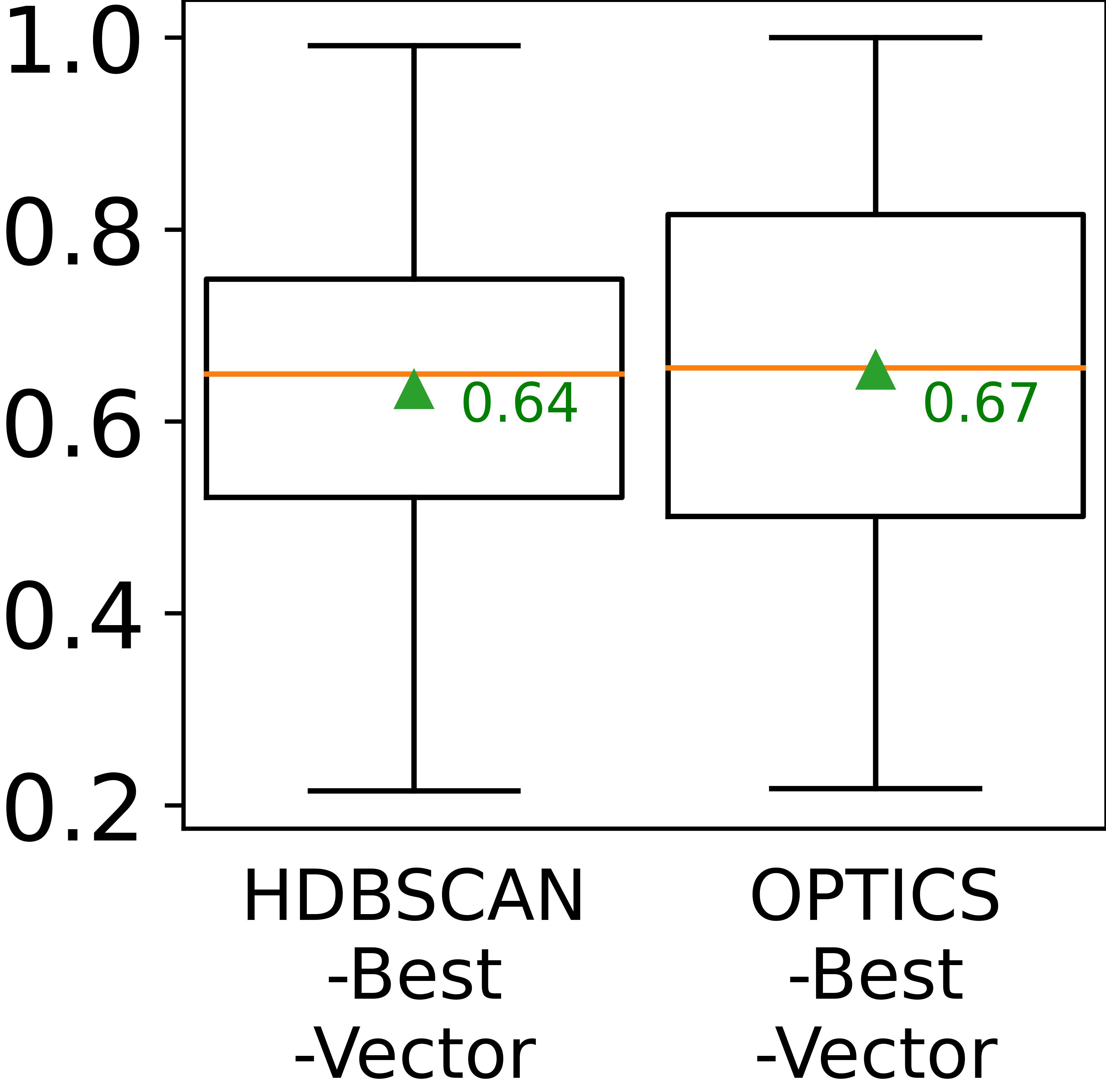}
    \includegraphics[width=0.21\textwidth]{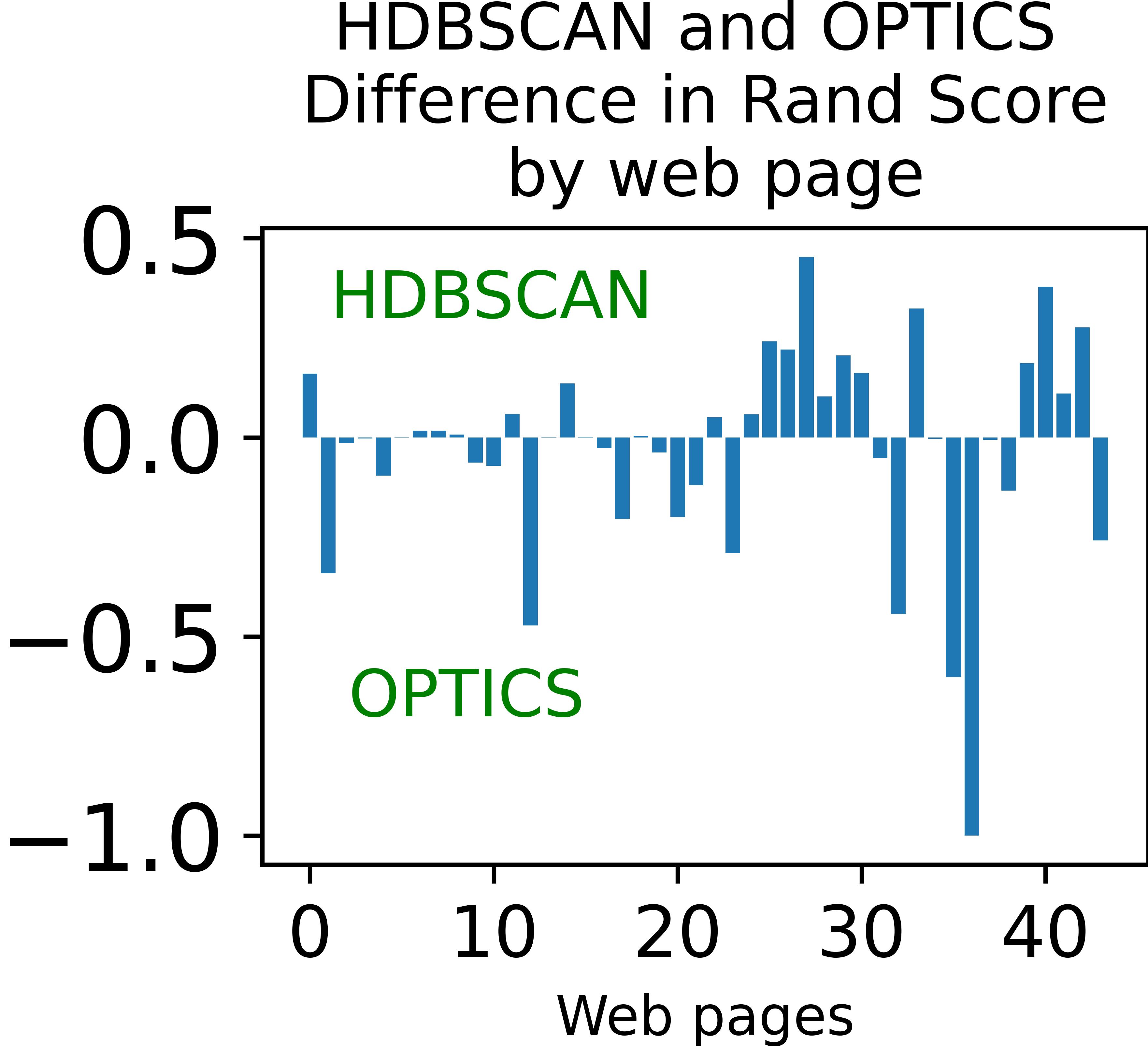}
    \vspace{-0.1in}
    \caption{Clustering Algorithms Performance}
    \label{fig:top-performers-algs-pie-charts}
\end{figure}

\begin{table}[]
    \caption{Testing Vector Performance and Noise Evaluations}
    \label{tab:results-from-experiments}
    \vspace{-0.1in}
    \centering
    \small
    \begin{tabular}{|c|c|c|c|c|}
         \hline
         & Testing Vector & Avg Rand & Cluster & Cluster \\
         & & & Count Dif. & Size Dif. \\ \hline
         1 & Tag Depth (TD) & 25.08\%+-20.97 & 48.61\%+-48 & 238.37\%+-201 \\ \hline
         2 & Div Id (DI) & 54.93\%+-24.16 & 21.01\%+-19 & 529.83\%+-538 \\ \hline
         3 & Data Id (DiD) & 37.27\%+-30.07 & 42.04\%+-49 & 346.06\%+-447 \\ \hline
        4 & Tag Group (TG) & 47.57\%+-23.57 & 33.9\%+-31 & 323.22\%+-283 \\ \hline
        5 & Cartesian (X,Y) & 20.39\%+-24.05 & 31.83\%+-28 & 306.68\%+-226 \\ \hline
         6 & True Cart. (TX,TY) & 18.35\%+-22.79 & 31.92\%+-30 & 319.63\%+-252 \\ \hline
        7 & TD-DI & 54.19\%+-23.87 & 20.3\%+-17 & 539.45\%+-541 \\ \hline
        8 & X-Y-TD-DI & 51.36\%+-23.81 & 20.39\%+-18 & 544.33\%+-553 \\ \hline
        9 & X-Y-TD-DI-TX-TY & 48.33\%+-22.93 & 20.65\%+-19 & 539.65\%+-549 \\ \hline
        10 & X-Y-TD-DI-DId-TG & 52.63\%+-25.25 & 20.47\%+-18 & 535.1\%+-539 \\ \hline
        11 & TD-DI-DId-TG & 52.82\%+-25.42 & 20.38\%+-18 & 534.74\%+-536 \\ \hline
        12 & TD-DI-TG & 53.96\%+-24.2 & 20.25\%+-17 & 539.62\%+-542 \\ \hline
        13 & All & 51.54\%+-24.52 & 20.22\%+-18 & 542.52\%+-549 \\ \hline
    \end{tabular}
\end{table}
\textbf{Clustering Algorithms.} We first evaluate the performance of OPTICS and HDBSCAN when the best-performing coordinate vectors are used. Fig.~\ref{fig:top-performers-algs-pie-charts} (left) shows that OPTICS performs slightly better on average with a rand score of 67\%, while HDBSCAN achieves 64\%. Hence, we utilize OPTICS for the rest of our evaluation. Digging deeper, Fig.~\ref{fig:top-performers-algs-pie-charts} (right) shows the difference in rand score of HDBSCAN and OPTICS per webpage. We find that on some pages, HDBSCAN performs better (+0.46 score diff), on others, OPTICS performs better (-0.9 score diff), and on some, both algorithms perform exactly the same (0 diff). We note an average performance gain of around 20\% for the entire dataset.

\begin{figure}[]
\centering
        \begin{subfigure}[b]{0.42\textwidth}
            \includegraphics[scale=0.22]{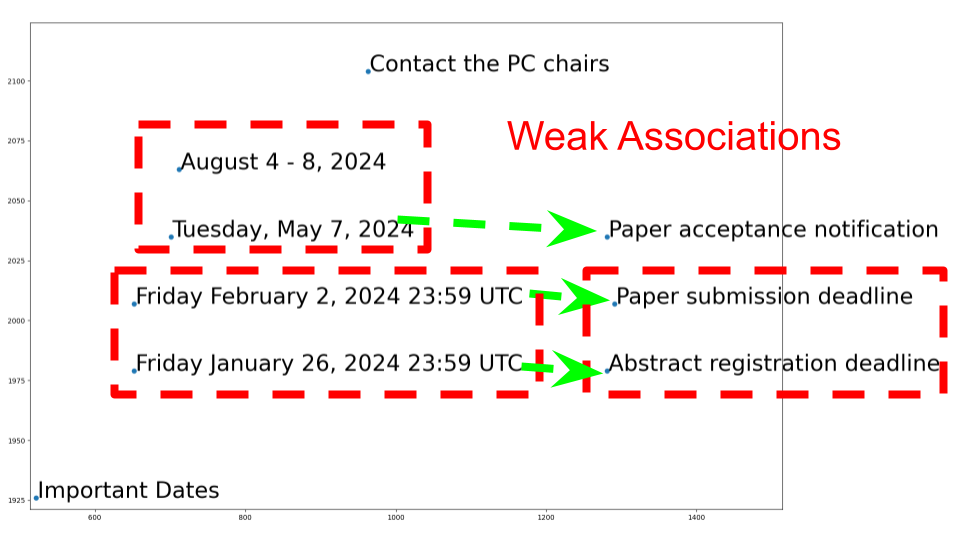}
            \caption{\emph{Example of Association with Visual Coordinates X-Y.}}
            \label{fig:example-visual}
        \end{subfigure}

        \begin{subfigure}[b]{0.42\textwidth}
            \includegraphics[scale=0.22]{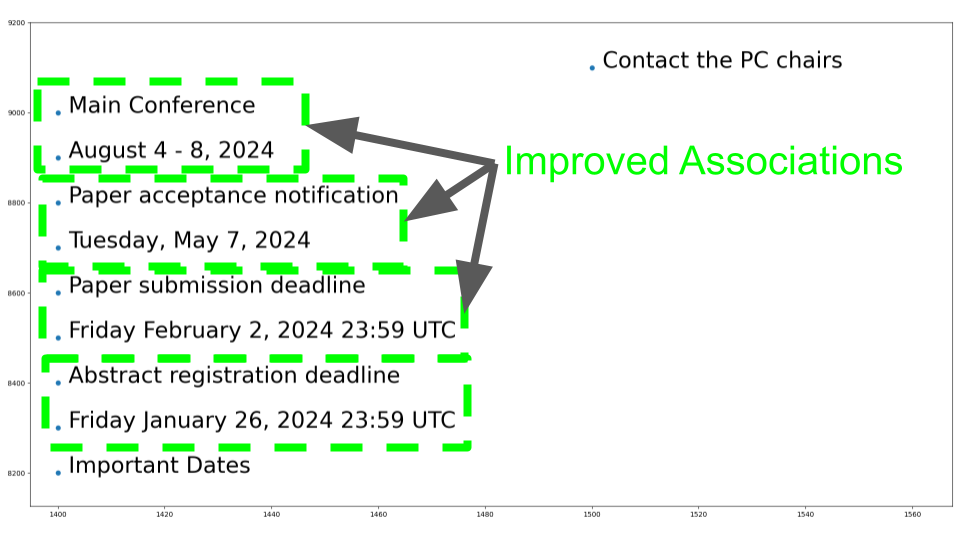}
            \caption{Example of Association with DOM Coordinates TD-DI.}
            \label{fig:example-dom}
        \end{subfigure}
    \caption{Example of DOM Improved Associations.}
    \label{fig:example-visual-vs-dom}
\end{figure}

\textbf{Vector Coordinates.} Table~\ref{tab:results-from-experiments} summarizes the performance of the different standalone and variable composition vectors using the OPTICS algorithm. From this table, we can notice that none of the vector coordinates can achieve a very high average Rand score; on average, each vector coordinate achieves approximately 50\%.
Notably, the visual coordinates $X$, $Y$ (row 5) and $TX$, $TY$ (row 6) performed markedly worse than the DOM-only coordinates (rows 1-4,7-13), achieving half of the Rand score. This is surprising, given the prevalence of such coordinates in existing work. We postulate that this performance is likely when the visual coordinates fail to capture logical associations. However, DOM coordinates are less susceptible to being impacted. An example of this resiliency is shown in Fig.~\ref{fig:example-visual-vs-dom}, where the visual coordinates create incorrect associations between items that are physically close together on the page, but not logically related. In contrast, the DOM coordinates in the example correct these associations. 

\begin{figure}[h]
    \centering
    \includegraphics[width=0.3\textwidth]{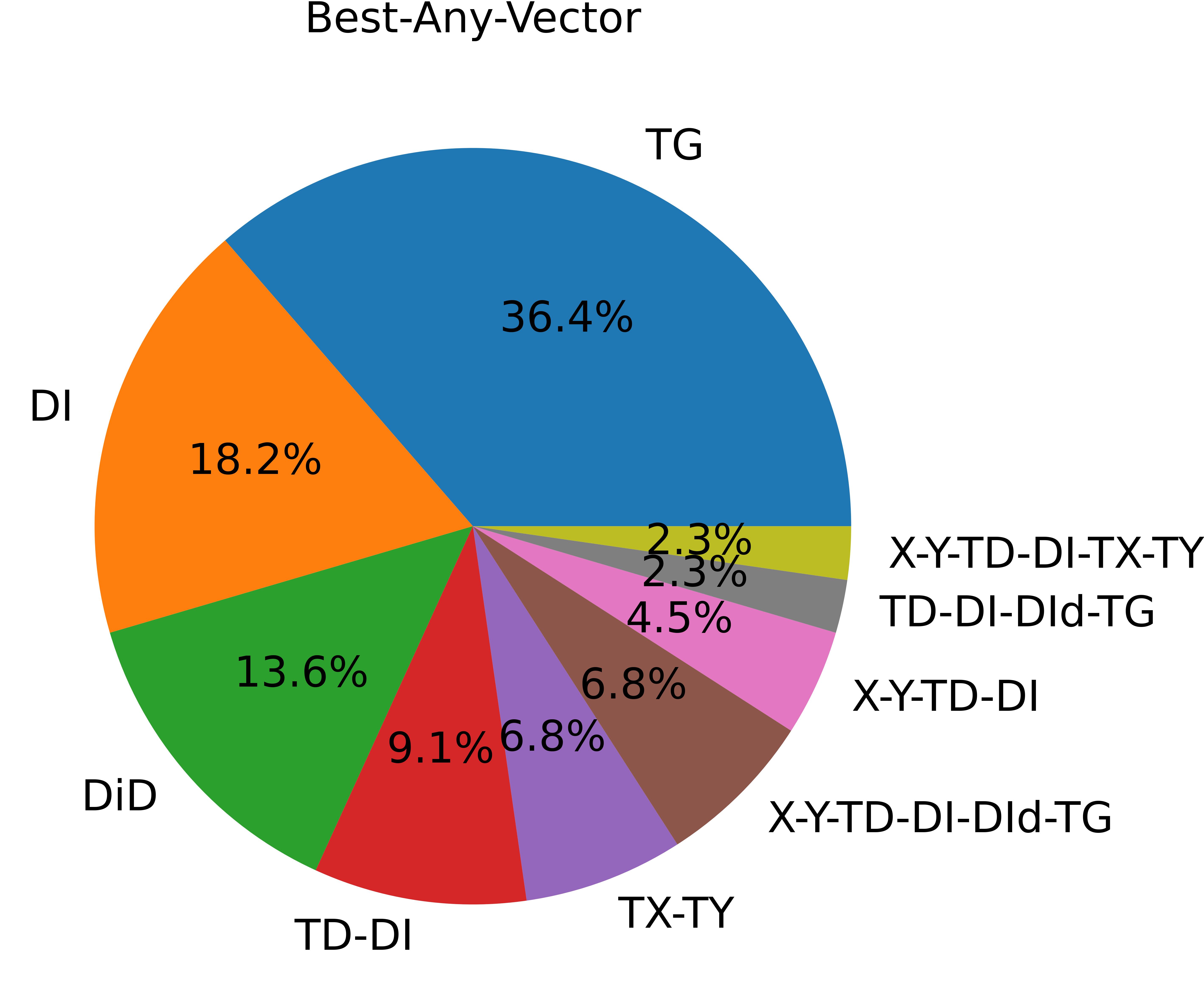}
    \vspace{-0.1in}
    \caption{Top Performing Vectors}
    \label{fig:top-performers-pie-charts}
\end{figure}

As with the clustering algorithms, no coordinate vector stands out. However, when looking at page-specific results, we find that some vectors perform exceptionally well or extremely poorly on.
Fig.~\ref{fig:top-performers-pie-charts} shows the Best-Any-Vector breakdowns for the different vector coordinates. We can notice that the DOM coordinates (TG, DI, and DiD) have the highest Rand score in 68.2\% of the web pages, while the visual coordinates $X$, $Y$ and $TX$, $TY$ show a minimal presence as top performers. This provides additional evidence that visual coordinates are less effective compared to DOM coordinates.
If the best vector is always selected (among single vectors and multi-vectors), the average Rand score can see a boost of 2\% on our dataset. 
 
\begin{figure}[t]
    \centering
    \includegraphics[width=0.2\textwidth]{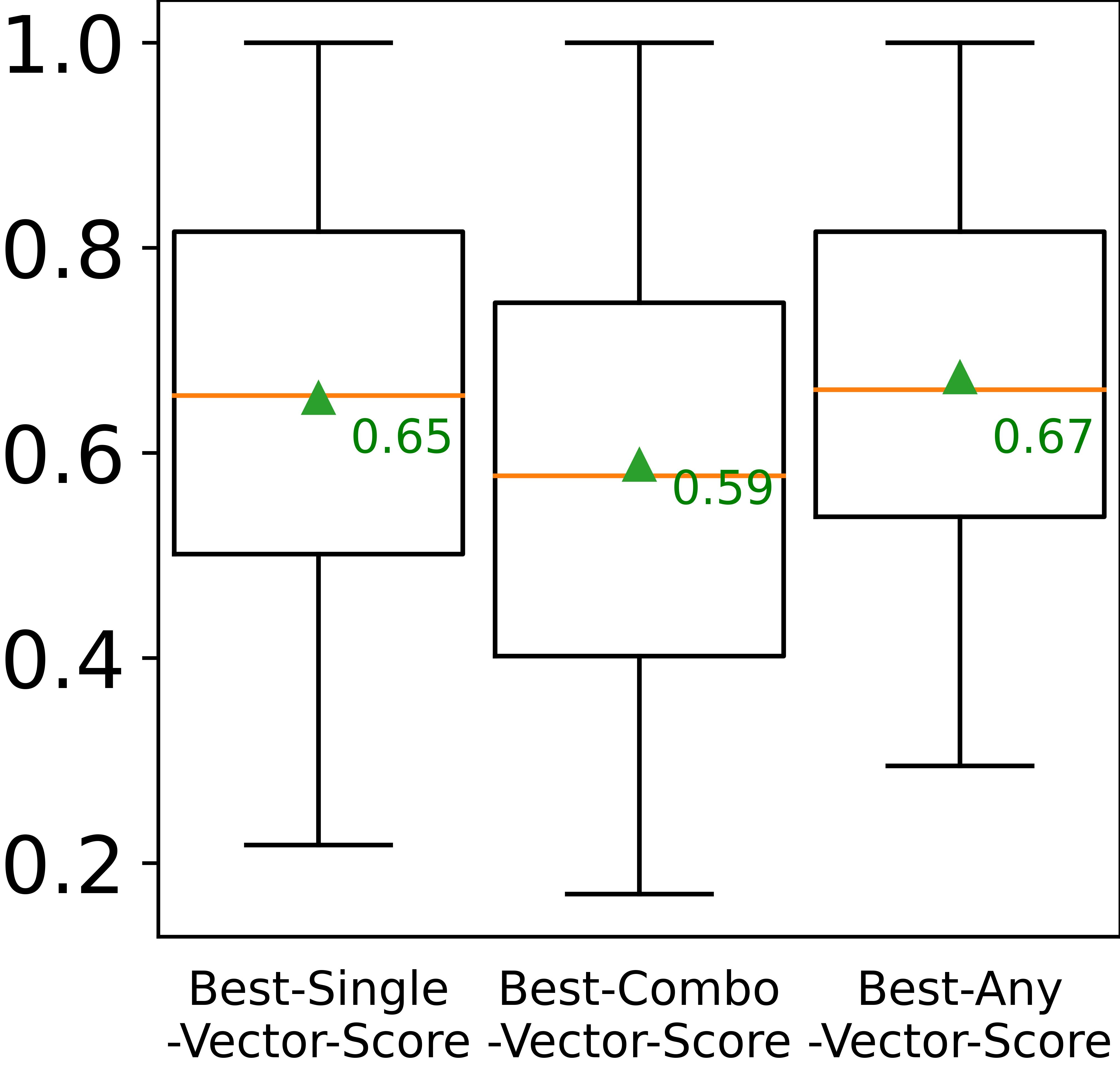}
    \vspace{-0.1in}
    \caption{Best Page-Vector Performance (OPTICS)}
    \label{fig:top-performers-boxplot-charts}
\end{figure}

In Fig.~\ref{fig:top-performers-boxplot-charts}, we compare the average Rand score if the best single coordinate vector, combined vector, and any vector are used for clustering. We find that the best single vector achieves 65.9\%, which is 7\% higher than the best combined vector. Moreover, if the best clustering algorithm is paired with the best performing vector (selected from both multi-vectors and single-vectors), the overall average Rand score can be boosted up to 74\%.

\textbf{Cluster Noise.} Additionally, as seen in Table~\ref{tab:results-from-experiments}, we find that the DOM coordinates have a ``noise'' character in terms of Cluster Count and Size.
These values are significantly large, which is expected as each item on the webpage is mapped into a cluster, there is a large number of items per webpage, and only subsets of a webpage are of relatable interest or part of a logical cluster, creating that significant difference. 
Mechanisms to reduce this noise are out of scope for this paper and fall under cluster coherence approaches that reason over the data held within a cluster. Our work would serve as input to these cluster-coherence approaches.

\section{Conclusion}
In this work, we explored the composition and construction of page segmentation vectors. We found that while there is no one-size-fits-all vector, generally speaking, DOM informed coordinates perform better than the traditional visual coordinates. Further, simple vectors composed of single coordinates tend to be the higher performers over more complex multi-coordinate vectors. 
Taken together this challenges the traditional conception of what makes a effective segmentation vector and emphasizes the need for greater evaluation and considering for such work. 

Additionally, we find that the diversity of web pages require a greater granularity of solutions for this type of segmentation. When able to match the right vector and algorithm to the right web page, the segmentation results greatly improve. This means that future mechanisms should consider a diverse approach or classification oriented approach for selecting the right segmentation.

Further, we note that such approaches still generate a significant amount of noise, but can be refined with additional coherence and refinement mechanisms with greater reasoning outside the scope of this paper.

Altogether our work provides a robust evaluation that can aid in the data extraction of web pages for emerging data hungry technologies such as LLMs, data aggregators, and such systems.

\bibliographystyle{ACM-Reference-Format}
\bibliography{refs.bib}


\end{document}